\documentclass[twocolumn,showpacs,preprintnumbers,amsmath,amssymb,pre]{revtex4}
\usepackage{graphicx}% Include figure files
\usepackage{dcolumn}% Align table columns on decimal point
\usepackage{bm}% bold math
\begin{document}
\title{Colloidal hard-rod fluids near geometrically structured substrates}
\author{L. Harnau$^{1,2}$, F. Penna$^3$, and S. Dietrich$^{1,2}$}
\affiliation{
         $^1$Max-Planck-Institut f\"ur Metallforschung,  
         Heisenbergstr.\ 3, D-70569 Stuttgart, Germany, 
         \\
         $^2$Institut f\"ur Theoretische und Angewandte Physik, 
         Universit\"at Stuttgart, 
         Pfaffenwaldring 57, 
         D-70569 Stuttgart, Germany\\
         $^3$Departamento de Fisica Te\'orica de la Materia Condensada,
         Universidad Aut\'onoma de Madrid, 
         E-28049 Madrid, Spain}
\date{\today}
\begin{abstract}
Density functional theory is used to study colloidal hard-rod fluids near 
an individual right-angled wedge or edge as well as near a hard wall which is 
periodically patterned with rectangular barriers. The Zwanzig model, in which the 
orientations of the rods are restricted to three orthogonal orientations but their 
positions can vary continuously, is analyzed by numerical minimization of the grand 
potential. Density and orientational order profiles, excess adsorptions, as well as 
surface and line tensions are determined. The calculations exhibit an enrichment 
[depletion] of rods lying parallel and close to the corner of the wedge [edge]. 
For the fluid near the geometrically patterned wall, complete wetting of the 
wall -- isotropic liquid interface by a nematic film occurs as a two-stage process 
in which first the nematic phase fills the space between the barriers until an almost 
planar isotropic -- nematic liquid interface has formed separating the higher-density 
nematic fluid in the space between the barriers from the lower-density isotropic 
bulk fluid. In the second stage a nematic film of diverging film thickness develops 
upon approaching bulk isotropic -- nematic coexistence. 
\end{abstract}

\pacs{61.30.Gd, 61.20.-p, 82.70.Dd}
\maketitle

\section{Introduction}
There is growing interest in properties of suspensions 
of colloidal particles near structured  walls because of useful applications 
such as selective deposition of particles \cite{dins:98} and controlled 
growth of colloidal crystals \cite{yin:02}. While experimental 
\cite{dins:98,yin:02}, theoretical \cite{kino:02a,kino:02b,bryk:03,cast:03}, 
and computer simulation \cite{shoe:97,boda:99,dies:00,scho:02} studies have 
been devoted to the understanding of the behavior of spherical colloidal 
particles near  geometrically structured substrates, suspensions of rodlike
colloidal particles in contact with such substrates have not been investigated 
yet, despite the importance of rodlike colloids for both biological and materials 
application. From a theoretical point of view, as compared with fluids consisting 
of spherical particles the study of rods is more difficult because of the 
additional orientational degrees of freedom. 

\begin{figure}[t]
\vspace*{-0.8cm}
\includegraphics[width=8.5cm,bb=105 0 750 640]{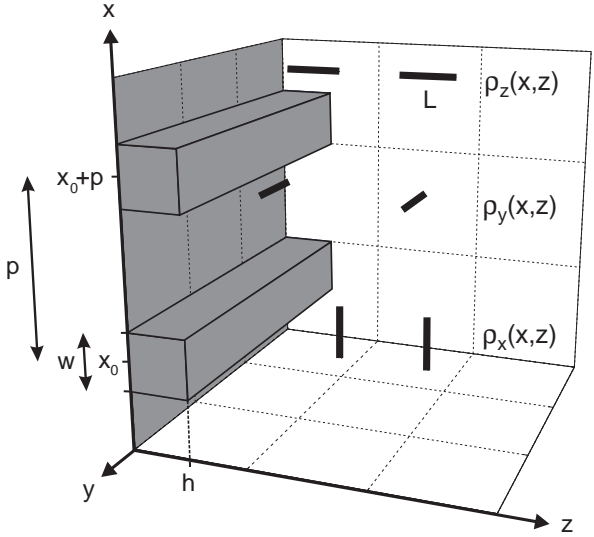}
\vspace*{0.5cm}
\caption{The system under consideration consists of a fluid of thin hard rods 
of length $L$ and thickness $D \ll L$ in contact with a hard substrate (grey).
The surface $z_s(x)$ of the substrate wall exhibits a periodic pattern with 
period $p$ consisting of rectangular blocks of width $w$ and height $h$. 
The density profiles for the centers of the rods with an orientation of the normal 
along their main axis of symmetry parallel to the orthogonal unit vectors of the  
reference frame are denoted as $\rho_x(x,z)$,  $\rho_y(x,z)$, and $\rho_z(x,z)$.
The system is spatially invariant in $y$-direction. In accordance with the 
discreteness of the orientational degrees of freedom 
$\rho_z(x,z<z_s(x)+L/2)=0$, 
$\rho_x(x_0+np-(w+L)/2<x<x_0+np+(w+L)/2,z<h)=0$ with $n \in \mathbb{Z}$, 
and $\rho_{x,y}(x,z<z_s(x))=0$ where $z_s(x_0+np-w/2<x<x_0+np+w/2)=h$ and 
$z_s(x)=0$ otherwise. The value of $x_0$ is arbitrary.}
\label{fig1}
\end{figure}

Here we study hard-rod fluids near geometrically structured walls 
within the Zwanzig approximation \cite{zwan:63}. In this model the allowed 
orientations of the rods are restricted to three mutually perpendicular 
orientations rather than a continuous range of orientations in space 
(see Fig.~\ref{fig1}); the positions of the rod centers are continuous variables.
The advantage of this model is that the difficult 
determination of spatially inhomogeneous density and orientational order
profiles becomes feasible allowing one to study various aspects of hard-rod 
fluids near structured walls in detail. On the basis of recent theoretical 
studies on fluids of hard rods near planar hard walls 
\cite{roij:00,roij:00a,harn:02c}, we expect to find results which remain 
qualitatively correct even in the absence of the restriction to discrete 
orientational directions. In Sec. II we describe the density functional theory 
which is used to analyze a hard-rod fluid in contact with an individual 
right-angled wedge or edge (Sec. III) or with a periodically patterned wall 
(Sec. IV).
\section{Model and density functional theory}  
We consider an inhomogeneous fluid consisting of hard rods of length $L$ and 
diameter $D$. The number density of the centers 
of mass of the rods at point ${\bf r}$ with orientation $\omega=(\theta,\phi)$ of 
the normal along their main axis of symmetry is denoted as $\rho({\bf r},\omega)$. 
The equilibrium density profile of the inhomogeneous 
liquid under the influence of an external field $V({\bf r},\omega)$ minimizes 
the grand potential functional
\begin{eqnarray} \label{eq1}
\Omega[\rho({\bf r},\omega)]\!\!\!&=&\!\!\!\int d^3r\,d\omega\,\rho({\bf r},\omega)
\left[k_BT\left(\ln[4\pi\Lambda^3\rho({\bf r},\omega)]-1\right)\right.\nonumber
\\ &&-\left.-\mu+V({\bf r},\omega)\right]+ 
F_{ex}[\rho({\bf r},\omega)]\,,\nonumber
\\
\end{eqnarray}
where $\Lambda$ is the thermal de Broglie wavelength and $\mu$ is the 
chemical potential. Within the Onsager second virial approximation the free 
energy functional $F_{ex}[\rho({\bf r},\omega)]$ in excess of the ideal gas 
contribution reads \cite{onsa:49}
\begin{eqnarray} \label{eq2}
F_{ex}[\rho({\bf r},\omega)]&=&-\frac{k_BT}{2}\int d^3r_1\,d\omega_1
\,d^3r_2\,d\omega_2\,\rho({\bf r}_1,\omega_1)\nonumber
\\&&\times \rho({\bf r}_2,\omega_2)f({\bf r}_1,{\bf r}_2,\omega_1,\omega_2)\,,
\end{eqnarray}
where $f({\bf r}_1,{\bf r}_2,\omega_1,\omega_2)$ is the Mayer function of 
the interaction potential between two rods. The Mayer function equals 
$-1$ if the rods overlap and is zero otherwise.

In the present application of density functional theory we concentrate on the 
ordering effects induced by surfaces geometrically structured such that the 
resulting $\rho({\bf r},\omega)$ depends on two spatial coordinates.
For the model system displayed in Fig.~\ref{fig1}, apart from the possibility of
surface freezing at high densities, non-uniformities of the density occur only in the 
\mbox{$x$-$z$} plane, so that $\rho({\bf r},\omega)=\rho(x,z,\theta,\phi)$. 
Minimization of $\Omega$ with respect to 
$\rho(x,z,\theta,\phi)$ leads to the following Euler-Langrange equation:
\begin{eqnarray} \label{eq3}
&&k_BT\ln[4\pi\Lambda^3\rho(x,z,\theta,\phi)]=\mu- V(x,z,\theta,\phi)\nonumber
\\&&+
\int dx_1\,dz_1\,\int\limits_0^{2\pi} d\phi_1\,\int\limits_0^{\pi} d\theta_1\,
\sin\theta_1\,\rho(x_1,z_1,\theta_1,\phi_1)\nonumber
\\&&\times f(x,z,\theta,\phi,x_1,z_1,\theta_1,\phi_1)k_BT\,.
\end{eqnarray}
This equation can be solved numerically for a given chemical potential 
$\mu$ and a given external field $V(x,z,\theta,\phi)$. For computational 
purposes, the density profile has to be specified on a sufficiently fine
four-dimensional $(x,z,\theta,\phi)$ grid. In order to reduce this computational 
effort we use the Zwanzig model for rods \cite{zwan:63}. Within the 
Zwanzig model the rods are represented by rectangular blocks of size 
$L\times D\times D$. The positions of the center of mass vary continuously, while the 
allowed orientations of the normal of each rod are restricted to directions 
parallel to the $x$, $y$, and $z$ axis (see Fig.~\ref{fig1}). Using the notation 
$\alpha_x(x,z)=\alpha(x,z,\theta=\pi/2,\phi=0)$, 
$\alpha_y(x,z)=\alpha(x,z,\theta=\pi/2,\phi=\pi/2)$, and
$\alpha_z(x,z)=\alpha(x,z,\theta=0,\phi=0)$ with $\alpha=\rho, V$,
the Euler-Langrange equations for a fluid consisting of thin Zwanzig rods 
($D/L\to 0$) can be written as
\begin{eqnarray} \label{eq4}
\ln[\Lambda^3\rho_x(x,z)]&=&(k_BT)^{-1}(\mu-V_x(x,z))\nonumber
\\&&-2D\int_{x-L/2}^{x+L/2}dx_1\,\int_{z-L/2}^{z+L/2}dz_1\,\rho_z(x_1,z_1)\nonumber
\\&&-2DL\int_{x-L/2}^{x+L/2}dx_1\,\rho_y(x_1,z)\,,
\\\ln[\Lambda^3\rho_y(x,z)]&=&(k_BT)^{-1}(\mu-V_y(x,z))\nonumber
\\&&-2DL\int_{z-L/2}^{z+L/2}dz_1\,\rho_z(x,z_1)\nonumber
\\&&-2DL\int_{x-L/2}^{x+L/2}dx_1\,\rho_x(x_1,z)\,, \label{eq5}
\end{eqnarray}
and
\begin{eqnarray} 
\ln[\Lambda^3\rho_z(x,z)]&=&(k_BT)^{-1}(\mu-V_z(x,z))\nonumber
\\&&-2D\int_{x-L/2}^{x+L/2}dx_1\,\int_{z-L/2}^{z+L/2}dz_1\,\rho_x(x_1,z_1)\nonumber
\\&&-2DL\int_{z-L/2}^{z+L/2}dz_1\,\rho_y(x,z_1)\,,  \label{eq6}
\end{eqnarray} 
which allow for a straightforward iterative numerical computation on a 
two-dimensional $(x,z)$ grid. We note that the meaning of the terms on 
the right side of Eqs.~(\ref{eq4})-(\ref{eq6}) can easily be inferred from 
considering the various orientations of the rods (see Fig.~\ref{fig1}).
It is convenient to introduce the variable $\mu^\star=\mu-k_BT\ln(\Lambda^3/L^2D)$. 
In the following sections numerical data are given in terms of $\mu^\star$ and we 
drop the star in order to avoid a clumsy notation.

The bulk phase behavior of this model was studied a long time ago
by Zwanzig \cite{zwan:63}, who found a first-order isotropic -- nematic phase 
transition similar to Onsager's result for freely rotating rods \cite{onsa:49}. 
Recently, the Zwanzig model has been used to investigate the phase behavior of 
monodisperse \cite{roij:00,roij:00a} and binary \cite{harn:02c} rod fluids near a 
single planar hard wall and confined in slit pore. These calculations yield a 
wall-induced continuous surface transition from uniaxial to biaxial symmetry. 
Complete wetting of the wall -- isotropic liquid interface by a biaxial nematic 
film has been found. For the fluids confined by two parallel hard walls,
at large slit widths a first-order capillary nematization transition occurs, 
which terminates in a capillary critical point upon decreasing the 
slit width.

\section{Hard-rod fluid near a right-angled wedge and edge}
\begin{figure}[t]
\vspace*{-0.4cm}
\includegraphics[width=8.5cm,bb=100 0 750 610]{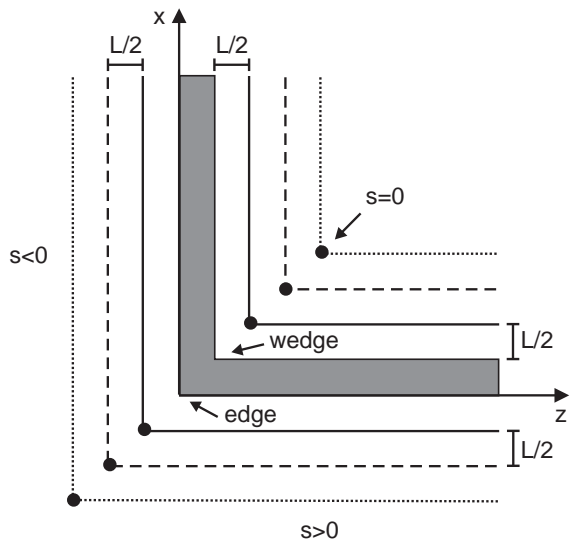}
\caption{Cross-section of an individual right-angled wedge and edge, respectively. 
The outer part of the shaded region forms an edge whereas the inner part forms a wedge. 
The geometrical structures are translationally invariant in the 
$y$-direction perpendicular to the plane of the figure. In Figs.~\ref{fig4} and 
\ref{fig5} density profiles of thin hard rods of length $L$ are shown along 
the paths indicated by the solid, dashed, and dotted lines which are parallel 
to the outer and inner surface, respectively. The paths are parameterized by $s$ 
such that for each line $s=0$ ($\bullet$) indicates the position of its corner 
at $x=z$. $\rho_y(x,z)$ is symmetric around $s=0$ and $\rho_x$ for $s\gtrless 0$ 
equals $\rho_z$ for $s\lessgtr 0$. Both for the wedge and edge the position 
of the corner is associated with $(x,z)=(0,0)$; in this figure both cases are 
superimposed.}
\label{fig2}
\end{figure}

Before studying the hard-rod fluid near the geometrically structured surface
shown in Fig.~\ref{fig1} it is instructive to analyze first the fluid around
an individual right-angled wedge or edge (see Fig.~\ref{fig2}). 
These 
simple geometrical structures constitute the building blocks of the structured 
surface displayed in Fig.~\ref{fig1}. An analysis of the size dependence leads 
to the following decomposition of the grand canonical potential functional of 
the fluid which in its bulk is taken to be in the isotropic phase:
\begin{eqnarray} \label{eq7}
\Omega[\rho(x,z)]=H_y\left[H_xH_z\omega_b+(H_x+H_z)\gamma_{wI}+
\tau_I(\alpha)\right]\,,
\end{eqnarray} 
where $\omega_b$ is the bulk grand canonical potential density,
$\gamma_{wI}$ is the wall -- isotropic liquid surface tension at a planar wall
and $\tau_I(\alpha)$ is the line tension of the isotropic liquid with 
$\alpha=\pi/2$ for the wedge and $\alpha=3\pi/2$ for the edge. The extension 
$H_\beta$ of the system in direction $\beta=x, y, z$
is defined as the length available to the rim of the particles at closest 
approach to the boundary. 
\begin{figure}[t]
\vspace*{-0.2cm}
\includegraphics[width=8cm]{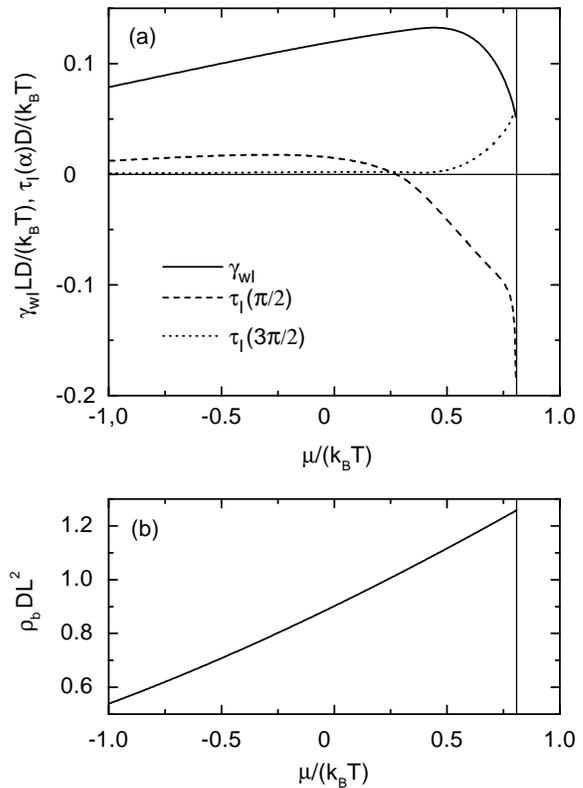}
\caption{(a) The wall -- isotropic liquid surface tension at a planar wall 
$\gamma_{wI}$ (solid line) together with the line tension at a right-angled 
wedge $\tau_I(\pi/2)$ (dashed line) and edge $\tau_I(3\pi/2)$ (dotted line) 
of a fluid consisting of thin rods of length $L$ and diameter $D$ ($D/L \to 0$)
as a function of the chemical potential. The vertical line marks the location 
of the chemical potential $\mu^{(IN)}/(k_BT)=0.8087227$ at bulk 
isotropic -- nematic coexistence. The wall -- isotropic liquid interface is 
completely wetted by a nematic film, i.e., 
$\gamma_{wI}=\gamma_{wN}+\gamma_{IN}=0.0498$ $k_BT/(LD)$ 
at $\mu^{(IN)}$, where $\gamma_{wN}$ is the  wall -- nematic liquid surface 
tension and $\gamma_{IN}$ is the isotropic -- nematic interfacial tension 
\cite{roij:00,roij:00a,harn:02c}. At $\mu^{(IN)}$, $\tau_I(3\pi/2)$ is slightly 
larger than $\gamma_{wI}L$ and $\tau_I(\pi/2)$ attains a finite value which 
becomes visible only at higher resolutions. 
$\tau_I(\pi/2)$ changes sign at $\mu/(k_BT)=0.27$. 
(b) The total particle number density $\rho_b=\rho(x,z\to \infty)$ 
[see Eq.~(\ref{eq11})] of the homogeneous and isotropic bulk fluid as a function 
of the chemical potential: $\mu/(k_BT)=ln(\rho_bDL^2)-\ln 3+4\rho_bDL^2/3$. 
At bulk isotropic -- nematic coexistence (marked by the vertical line) the density 
of the isotropic phase is given by $\rho_bDL^2=1.25822486$. This figure allows 
one to translate values for $\mu$ into $\rho_b$, which is experimentally 
accessible, and vice versa.}
\label{fig3}
\end{figure}
We restrict our analysis to chemical potentials 
$\mu$ smaller than the chemical potential $\mu^{(IN)}/(k_BT)=0.8087227$ at bulk 
isotropic -- nematic coexistence. Figure \ref{fig3} displays the surface tension 
$\gamma_{wI}$ as well as the line tensions $\tau_I(\pi/2)$ and $\tau_I(3\pi/2)$
as a function of the chemical potential. The steric interaction between the 
particles increases the surface tension with increasing chemical potential. 
On the other hand, the onset of the surface-induced nematic ordering of the 
particles leads to a decrease of the surface tension for larger chemical potentials. 
In the limit of large negative chemical potentials, i.e., for non-interacting 
particles, the wall -- isotropic liquid surface tension as well as the line 
tensions vanish. The line tension $\tau_I(\pi/2)$ for the fluid near a 
right-angled wedge  exhibits a change of sign with increasing 
chemical potential while $\tau_I(3\pi/2)>0$ for all values of $\mu$. We note 
that by construction $\tau_I(\pi)=0$. As a function of $\alpha$ the line 
tension $\tau_I(\alpha)$ corresponds to the work done per unit length, against 
the fluid, to change the dihedral angle from $\pi$ to some value $\alpha$
\cite{hend:02,hend:04}. The corresponding solvation torque 
is $t_I(\alpha)=-d \tau_I(\alpha)/d \alpha$.
\begin{figure}[t]
\vspace*{-0.9cm}
\includegraphics[width=8.5cm]{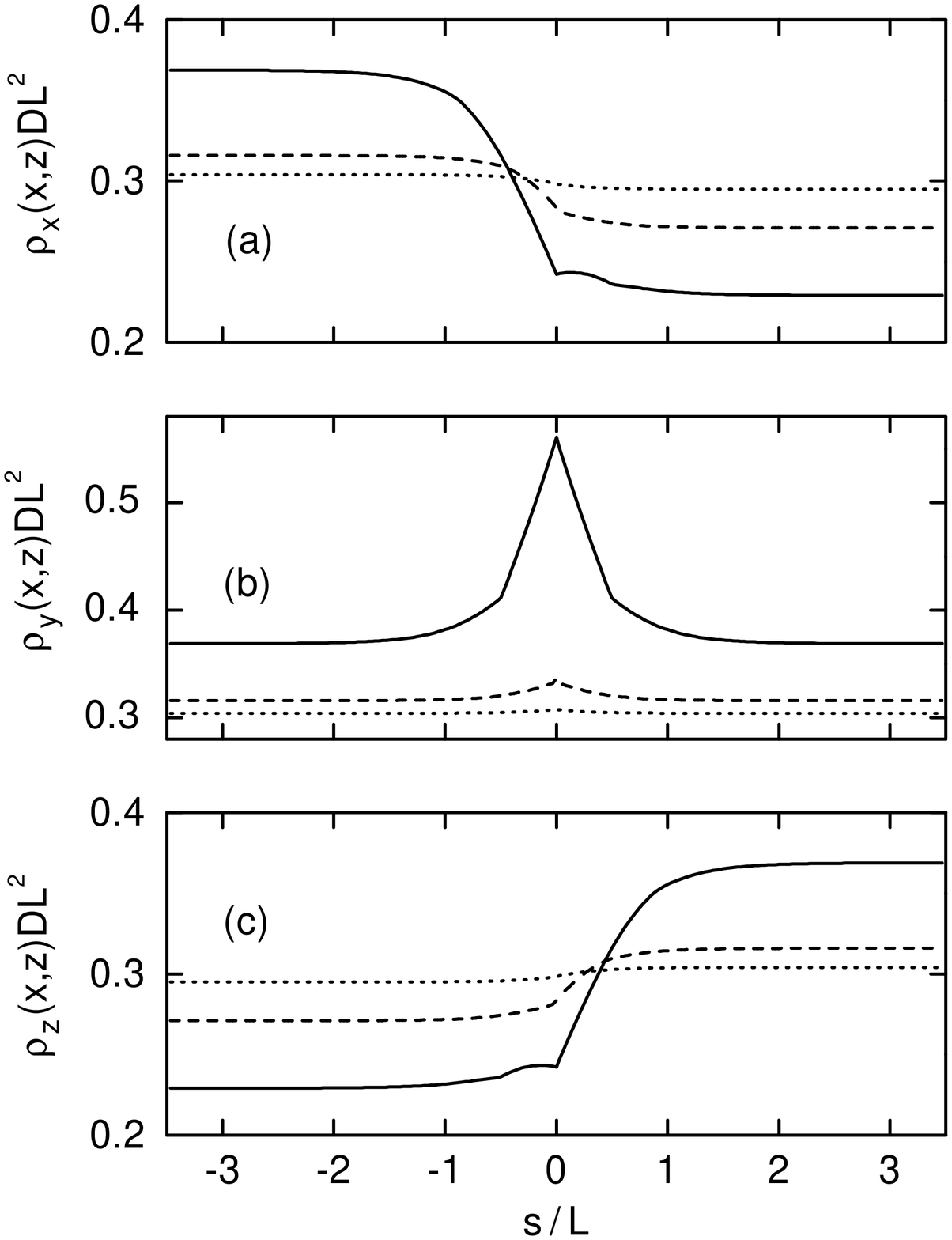}
\caption{The density profiles $\rho_x(x,z)$, $\rho_y(x,z)$, and $\rho_z(x,z)$
for thin hard rods of length $L$ and diameter $D$ ($D/L \to 0$) 
near a right-angled wedge evaluated along the paths specified in Fig.~\ref{fig2} 
[using the same line code, i.e., solid, dashed, and dotted lines]. 
The chemical potential is $\mu=0$. The solid curve in (a) [(c)] represents for 
$s\ge 0$ [$s\le 0$] the density profile along the line of closest contact,
i.e., rods oriented parallel to the $x$ [$z$] axis touch the wall with the rim.
$\rho_y$ is symmetric around $s=0$ and $\rho_x$ for $s\gtrless 0$ 
equals $\rho_z$ for $s\lessgtr 0$.}
\label{fig4}
\end{figure}

\begin{figure}[t]
\vspace*{-0.9cm}
\includegraphics[width=8.5cm]{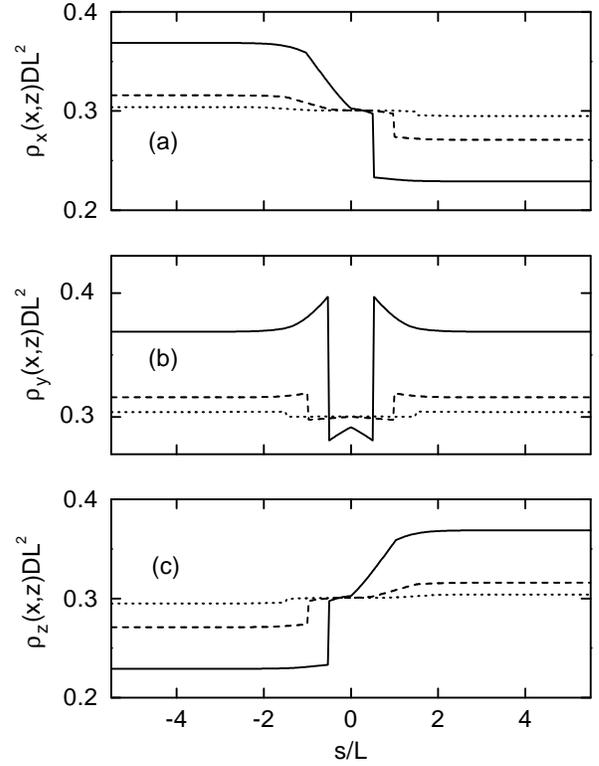}
\caption{The density profiles $\rho_x(x,z)$, $\rho_y(x,z)$, and $\rho_z(x,z)$
for thin hard rods of length $L$  and diameter $D$ ($D/L \to 0$) 
near a right-angled edge evaluated along the paths specified in Fig.~\ref{fig2} 
[using the same line code, i.e., solid, dashed, 
and dotted lines]. The chemical potential is $\mu=0$.
The solid curve in (a) [(c)] represents for $s\ge L/2$ [$s\le -L/2$]
the density profile along the line of closest contact, i.e., rods 
oriented parallel to the $x$ [$z$] axis touch the wall with the rim.
$\rho_y$ is symmetric around $s=0$ and $\rho_x$ for $s\gtrless 0$ 
equals $\rho_z$ for $s\lessgtr 0$.}
\label{fig5}
\end{figure}

Figures \ref{fig4} and \ref{fig5} show the density profiles at the right-angled
wedge and edge for a chemical potential $\mu=0$. The profiles are 
evaluated along lines parallel to the confining walls (see Fig.~\ref{fig2}), 
using a linear parametrization $s$ such that $s=0$ corresponds to the position 
of the corner. For $s\to \pm\infty$ the density profiles at a planar hard wall are 
recovered. As is apparent from Fig.~\ref{fig4} the density profile $\rho_y(x,z)$ 
exhibits a maximum at $s=0$ for the right-angled wedge. This maximum increases upon 
approaching the corner of the wedge. For a right-angled edge a significant depletion 
of rods lying parallel to the $y$ axis is found near the corner (see Fig.~\ref{fig5}). 
The sharp features of the density profiles in the presence of the right-angled edge 
are caused by the vanishing of $\rho_x(x\ge -L/2,z\ge 0)$ and 
$\rho_z(x \ge 0,z\ge -L/2)$ which is due to the presence of the impenetrably hard walls. 
Such cusps and discontinuities -- although less pronounced -- have already been found 
for the density profiles near a planar hard wall using the Zwanzig model 
\cite{roij:00,roij:00a,harn:02c},
while the interfacial profiles between demixed fluid phases, as calculated for 
the same model, exhibit neither cusps nor discontinuities \cite{bier:04}. 
We do not expect to observe discontinuities in the density profiles for freely 
rotating rods near an edge, similar to our findings for hard rods near a planar hard 
wall \cite{harn:02a} (see discussion below). Although the calculated density profiles 
presented in Figs.~\ref{fig4} and \ref{fig5} can only be considered to be of 
qualitative significance we expect that the main features, namely the enrichment 
[depletion] of rods lying parallel and close to the corner of a wedge [edge] as 
well as the asymmetry of the density profiles $\rho_x(x,z)$ and $\rho_z(x,z)$ with 
respect to the line $s=0$, remain valid for freely rotating rods. 

\begin{figure}[t]
\includegraphics[width=7.3cm,bb=100 0 750 610]{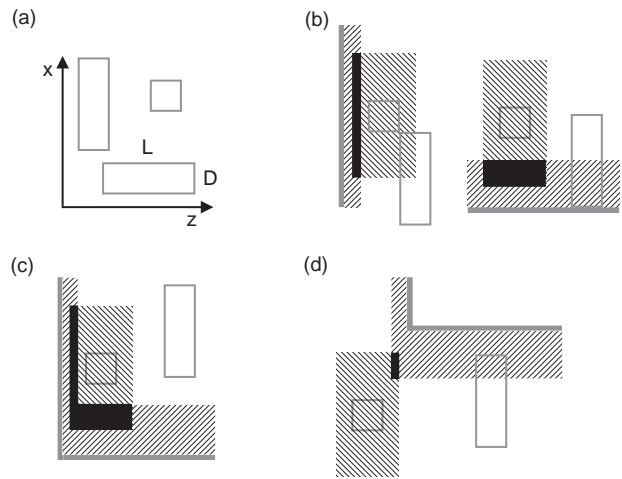}
\vspace*{0.5cm}
\caption{
Illustrations of the effective entropic interactions of hard rods with surfaces.
(a) Schematic side view of rectangular rods of length $L$ and 
thickness $D$ with orientations along the $x$, $y$, and $z$ 
axis, where the $y$-axis is perpendicular to the plane of the figure.
(b) Due to the steric interaction, the centers of mass of rods with $x$-direction
are excluded from the hatched regions surrounding the rods with $y$-direction 
(quadratic cross-section) and the walls (thick grey lines). Here one 
rod with $x$-direction touches a rod with $y$-direction (left) while another rod 
with $x$-direction fully touches the wall which is oriented parallel to the $z$-axis
(right). The rods with $x$-direction are viewed as depletion agents leading to 
an effective interaction between the rod with $y$-direction and the walls.
When rods with orientations along the y axis are sufficiently separated from 
each other and from the walls, the volume accessible to the rods with $x$-direction
is the total volume of the container minus the volume of the hatched regions. 
However, when a rod oriented parallel to the $y$-axis is close to a wall the 
volume accessible to the rods with $x$-direction increases by the excluded-volume
overlap region marked in black. The corresponding increase in entropy induces an 
effective attractive force between  
rods with an orientation parallel to the $y$-axis and the wall. (c) In a corner of a 
right-angled wedge (thick grey lines), the overlap volume (black region) is larger 
than the one on a flat wall leading to an enrichment of rods lying parallel and 
close to the corner of the wedge [see Fig.~\ref{fig4} (b)]. (d) Close to a 
right-angled edge (thick grey lines), the overlap volume is smaller than the one 
on a planar wall. Therefore the density of rods lying parallel and close to the 
edge is smaller than the density near a planar hard wall at the same chemical 
potential [see Fig.~\ref{fig5} (b)]. Similar considerations hold for the rods with 
$z$-direction acting as depletion agents on the rod with $y$-direction. The 
rods with $y$-direction are exposed to the superposition of both effective 
interactions.}
\label{fig6}
\end{figure}

In order to understand the structure of the density profiles displayed in 
Fig.~\ref{fig4} and \ref{fig5}, it is instructive to apply the idea of 
entropically driven forces (see, e.g., Ref. \cite{roth:00}) to the Zwanzig model. 
We consider rods of a given orientation along the x, y, and z axis  as belonging
to one of three ''species''. Such a three-component fluid maximizes its
entropy by maximizing the volume accessible per rod. 
Although there exist
only steric repulsions between pairs of particles, maximizing the entropy 
in the fluid mixture can lead to an effective entropic attraction between the rods 
and the walls. Figure \ref{fig6} (b) demonstrates that, when a rod of a given 
species approaches a planar wall (represented in grey), the total volume 
{\it available} to rods of the other species increases. This increases the total 
entropy of the mixture by an amount proportional to the size of the excluded-volume 
overlap region (represented in black) multiplied by the pressure. For a rod lying 
close and parallel to a right-angled wedge [edge] the corresponding excluded-volume 
overlap region is increased [decreased] (see Figs.~\ref{fig6} (c) and (d), 
respectively) leading to an enrichment [depletion] of such rods close to 
the corner of a wedge [edge]. The results may be interpreted in 
terms of a repulsive barrier of an effective potential repelling a rod, which 
is oriented parallel to the corner of an edge, approaching the edge from the side 
and practically preventing it from passing around the corner. On the other hand 
the effective potential acting on a rod which is oriented parallel to the 
corner of a wedge is ''pushing'' it into the corner. For a detailed analysis 
of these mechanisms acting on mixtures of hard spheres near edges and wedges 
see Ref. \cite{bryk:03} and in particular Figs. 5 - 7 therein.

The simple illustration in Fig.~\ref{fig6} (d) is also helpful for understanding 
the aforementioned discontinuities in the density profiles near the right-angled 
edge (see Fig.~\ref{fig5}). When a thin rod ($D/L \to 0$), which is oriented parallel 
to the edge, approaches the edge from the side, the excluded-volume overlap drops 
abruptly to zero  before the rod is passing around the corner. This causes the 
discontinuities in the density profiles along the paths specified in Fig.~\ref{fig2}. 
For freely rotating rods the corresponding excluded-volume overlap decreases 
smoothly to zero because of the huge number of differently oriented rods acting 
on the rod which is oriented parallel to the edge.

Finally, we briefly discuss the phase behavior of hard rods confined in a hard pore of 
square cross-section. The immediate consequence of the pore is that rods oriented 
perpendicular to the confining walls cannot approach closer than a center-of-mass 
distance $L/2$. There is a pronounced increase of the density of rods orientated 
parallel to the main axis of the pore in the corners of the pore because 
of the aforementioned effective entropic attraction. 
For sufficiently large cross-sections
of the pore, we observe coexistence between an isotropic phase and a capillary condensed 
nematic phase. The density profile of the capillary condensed nematic phase is 
characterized by a nematic phase throughout the pore, whereas the density profile of 
the coexisting phase decays toward an isotropic phase in the middle of the pore. For 
small pore cross-sections a sharp capillary nematization transition no longer occurs 
and is replaced by a steep but continuous filling upon increasing the chemical potential. 
For the same fluid confined in a slit pore the confinement effects are weaker. 
Thus, in the slit pore we observe capillary nematization at a higher chemical potential 
corresponding to a higher particle number density of the bulk fluid. However, the 
spatially averaged particle number density of the coexisting inhomogeneous isotropic 
phase in the slit pore is smaller than the corresponding one in the pore of square 
cross-section.

\section{Hard-rod fluid in contact with a periodically structured hard wall}
\begin{figure}[t]
\vspace*{-0.4cm}
\includegraphics[width=8.5cm]{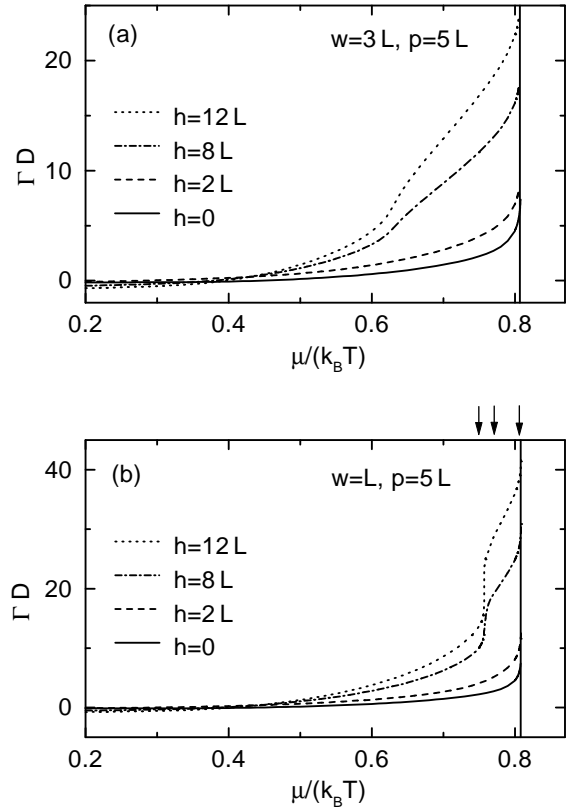}
\caption{The excess adsorption $\Gamma$ [see Eq.~(\ref{eq12})] of a fluid consisting 
of thin hard rods of length $L$ and diameter $D$ ($D/L \to 0$)
near a geometrically structured wall as shown in Fig.~\ref{fig1} for 
various barrier heights:
$h=0$ (solid curves); $h=2\,L$ (dashed curves);
$h=8\,L$ (dash-dotted curves); $h=12\, L$ (dotted curves).
The periodicity is $p=5\,L$ and the barrier width is $w=3\,L$ in (a) and $w=L$ in (b). 
The vertical lines mark the value of the chemical potential $\mu^{(IN)}/(k_BT)=0.8087227$
at bulk isotropic -- nematic coexistence. In all cases $\Gamma$ diverges logarithmically
as $\mu\to\mu^{(IN)}$. Figure \ref{fig8} exhibits density profiles for the 
system with $h=8\, L$ in (b) for the three chemical potentials 
$\mu=0.75$, $0.77$, $0.808$ $k_BT$ marked by arrows. 
The chemical potential $\mu\simeq 0.757$ $k_BT$ corresponding to the pronounced 
variation (no jump but steep increase) of $\Gamma$ for $h=12\, L$ in (b) turns out 
to agree with the chemical potential at the occurrence of the first-order capillary 
nematization transition of the same fluid confined in a corresponding slit pore of 
width $p-w=4\, L$. Neither the curves in (a) nor in (b) intersect at
a single point.}
\label{fig7}
\end{figure}

We now turn our attention to the properties of the hard-rod fluid in contact with 
the hard wall shown in Fig.~\ref{fig1}. The surface of the wall is periodically 
patterned with rectangular hard barriers of width $w$ and height $h$, 
where the periodicity is denoted by  $p$. We focus on the numerically determined 
orientationally averaged number density profile 
\begin{equation} \label{eq11}
\rho(x,z)=\rho_x(x,z)+\rho_y(x,z)+\rho_z(x,z)
\end{equation}
and the excess adsorption $\Gamma$ defined as
\begin{equation} \label{eq12}
\Gamma=\int dx\,dz\, [\rho(x,z)-\rho_b]\,,
\end{equation}
where $\rho_b=\rho(x,z\to\infty)$ is the total particle number density of the 
homogeneous bulk fluid. The volume $V=\int dx\,dy\,dz$ of the system is defined 
as the total volume of the container, i.e., the left boundary of the system 
displayed in Fig.~\ref{fig1} is taken to be the surface $z_s(x)$ of the substrate
wall which implies that the trenches between the barriers contribute to $V$. 
Figure \ref{fig7} displays $\Gamma$ for 
various values of the barrier height $h$ and two values of the barrier width $w$ at
a fixed periodicity $p$ of the surface pattern. For non-interacting rods
($\mu \to -\infty$), the calculated excess adsorption reveals a slight depletion 
close to the surface ($\Gamma<0$) because there is less space 
available to the rods in the presence of the impenetrably hard walls. For the 
same reason this depletion becomes more pronounced with increasing height of 
the barriers (i.e., increasing the actual exposed solid area). Upon increasing the 
chemical potential, the excess adsorption increases and exhibits a change of sign
because of the aforementioned entropic attraction between the rods and the 
surface [see Fig.~\ref{fig6}]. For small barrier heights, $\Gamma$ increases 
smoothly upon increasing the chemical potential, while a pronounced variation of 
the excess adsorption is found for large barrier heights at a chemical potential 
smaller than the chemical potential $\mu^{(IN)}$ at bulk isotropic -- nematic 
coexistence. Moreover, the calculation renders $\Gamma$ to diverge logarithmically 
as $\mu\to \mu^{(IN)}$. Near $\mu^{(IN)}$ the excess adsorption can be fitted by 
$\Gamma=A_1-A_2 \ln[(\mu^{(IN)}-\mu)/(k_BT)]$, with fit parameters $A_1$ and $A_2$,
where $A_2>0$ turns out to be independent of the surface pattern. 

The logarithmic divergence of $\Gamma$ is consistent with complete wetting of the 
wall - isotropic fluid interface by a nematic film in the absence of algebraically 
interaction potentials \cite{diet:88}. A similar behavior of the excess adsorption
close to $\mu^{(IN)}$ has been found for the same fluid near a planar hard wall 
\cite{roij:00,roij:00a,harn:02c}. 

\begin{figure}[t]
\vspace*{-1.5cm}
\mbox{
\hspace*{-.6cm}
\includegraphics[width=5.4cm]{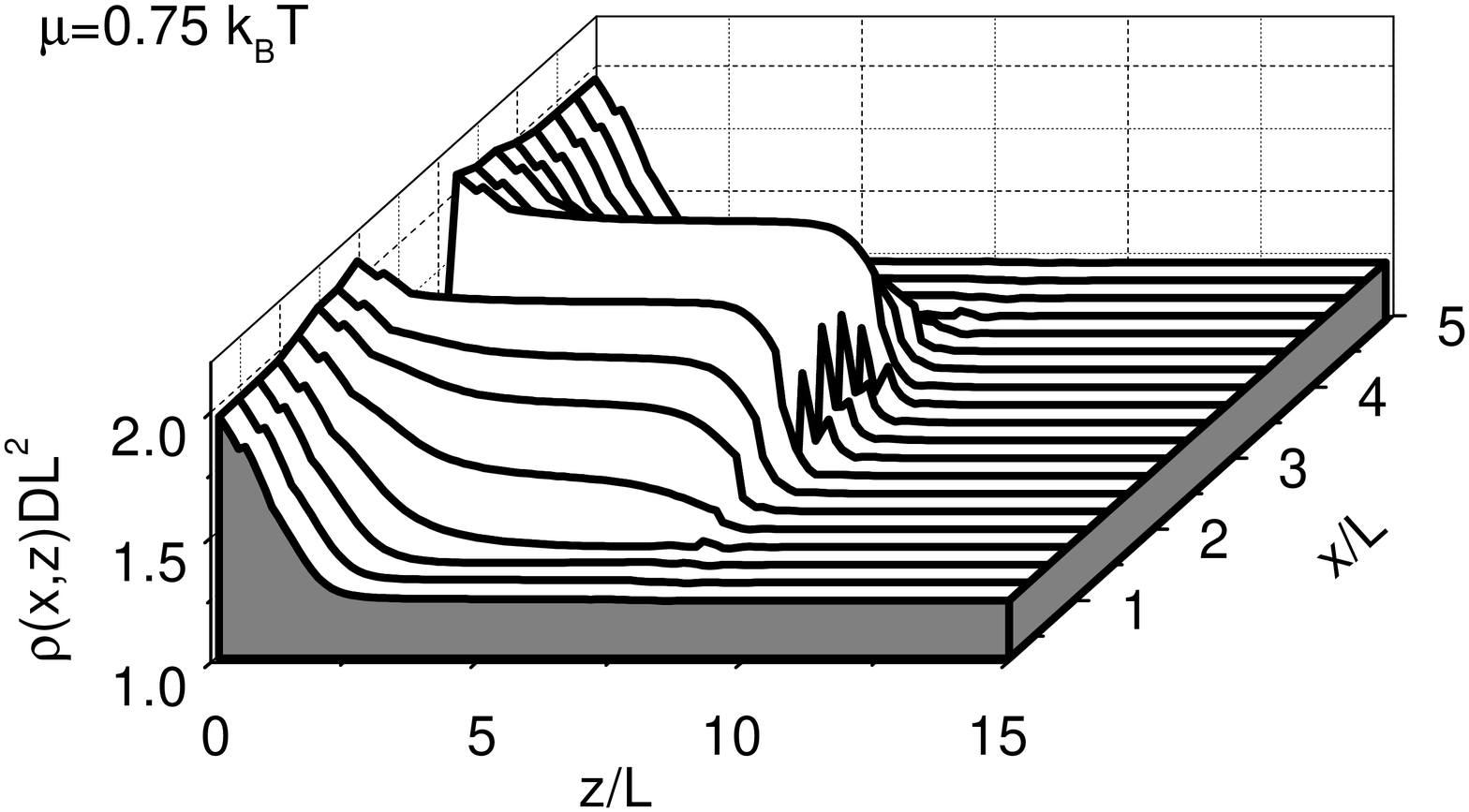}
\includegraphics[width=3.6cm]{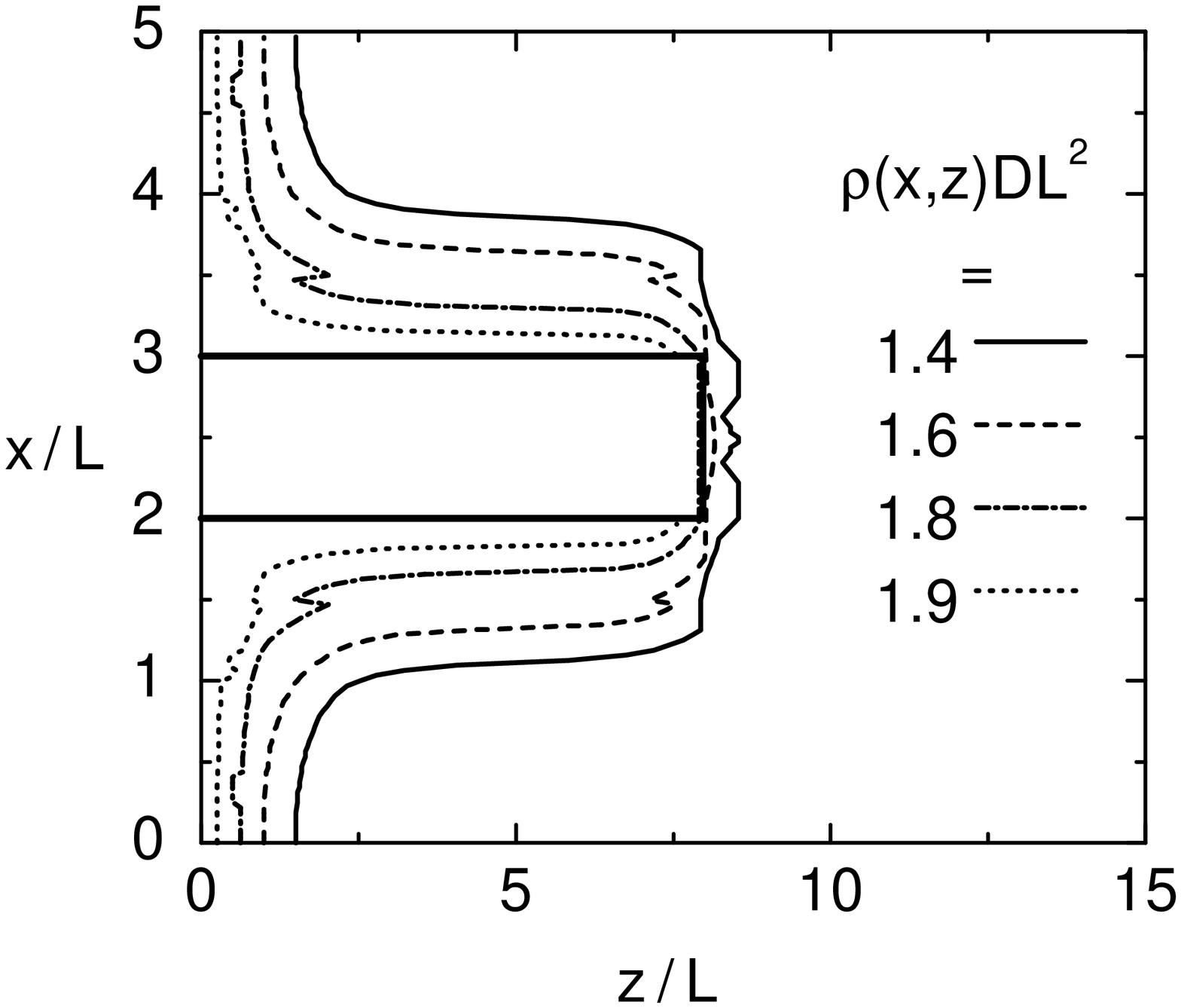}}
\\\vspace*{-1.3cm}
\mbox{
\hspace*{-.6cm}
\includegraphics[width=5.4cm]{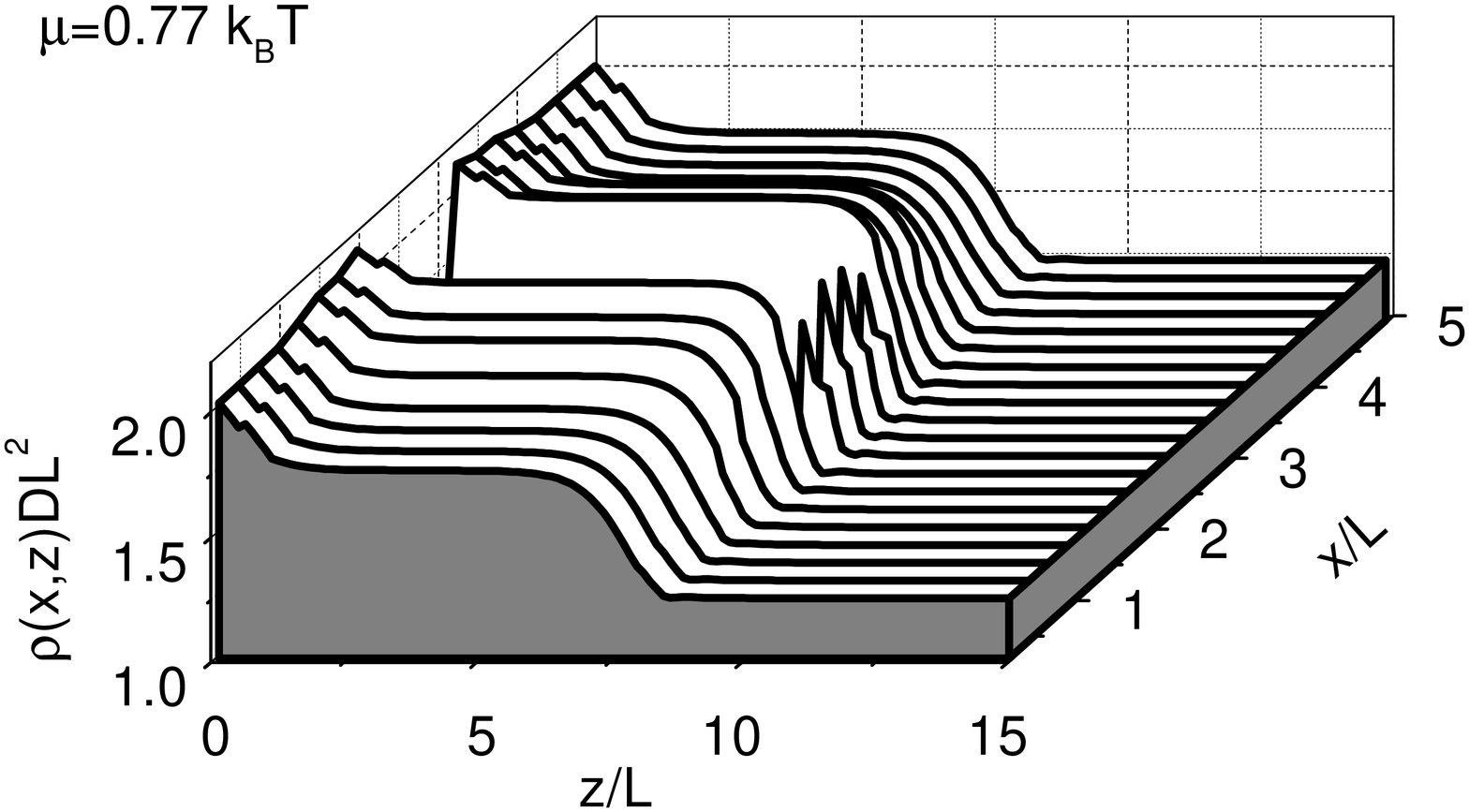}
\includegraphics[width=3.6cm]{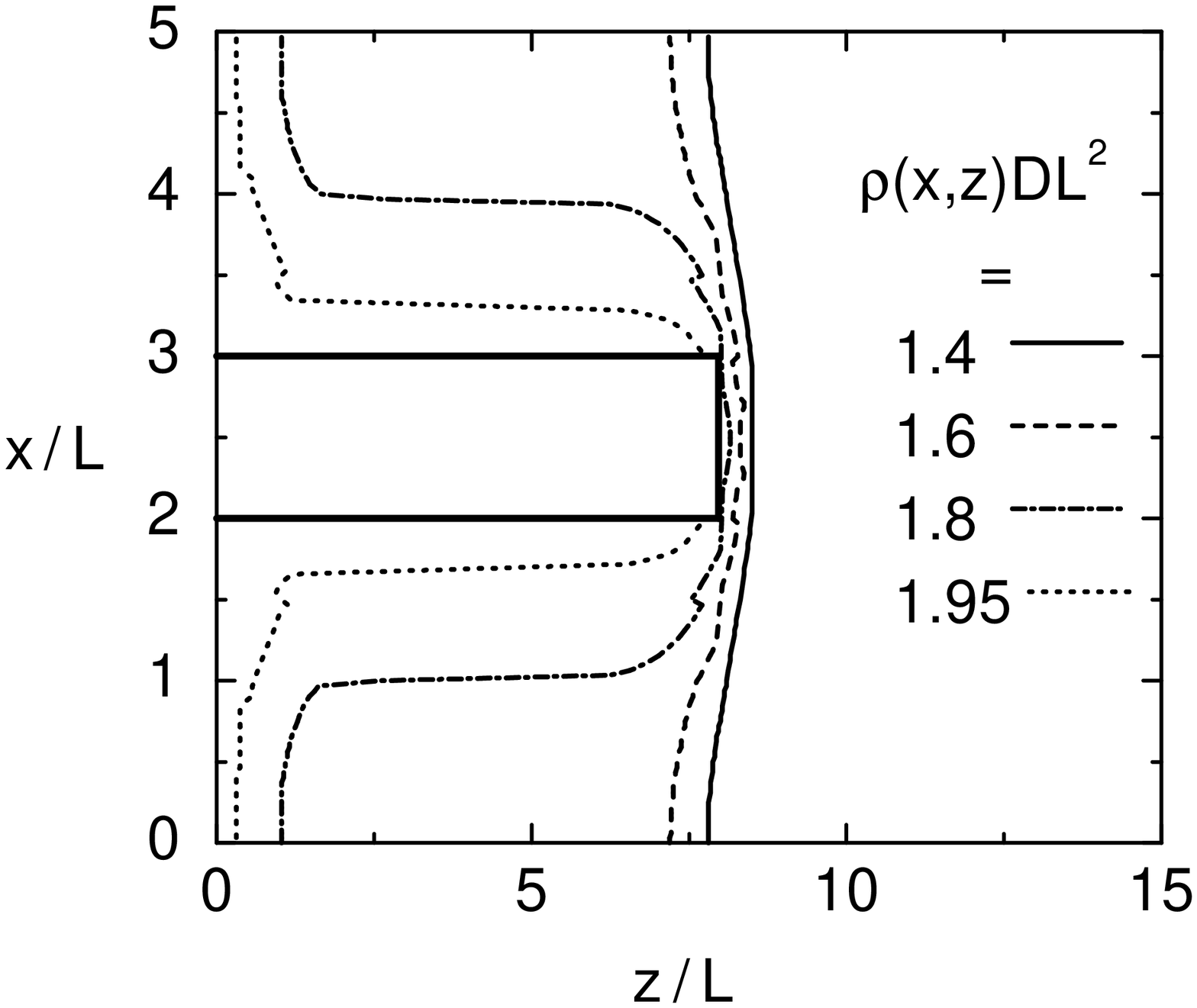}}
\\\vspace*{-1.3cm}
\mbox{
\hspace*{-.6cm}
\includegraphics[width=5.4cm]{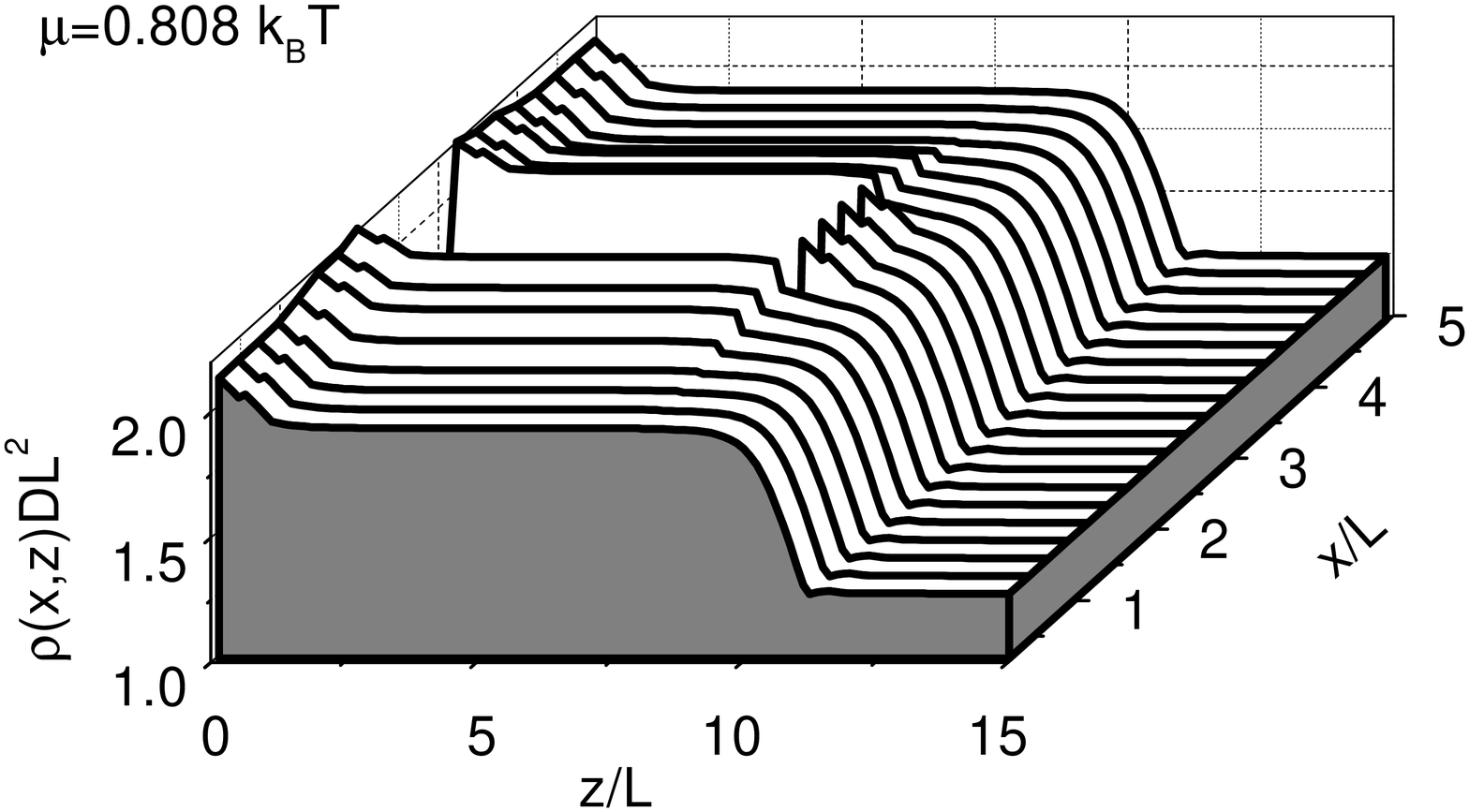}
\includegraphics[width=3.6cm]{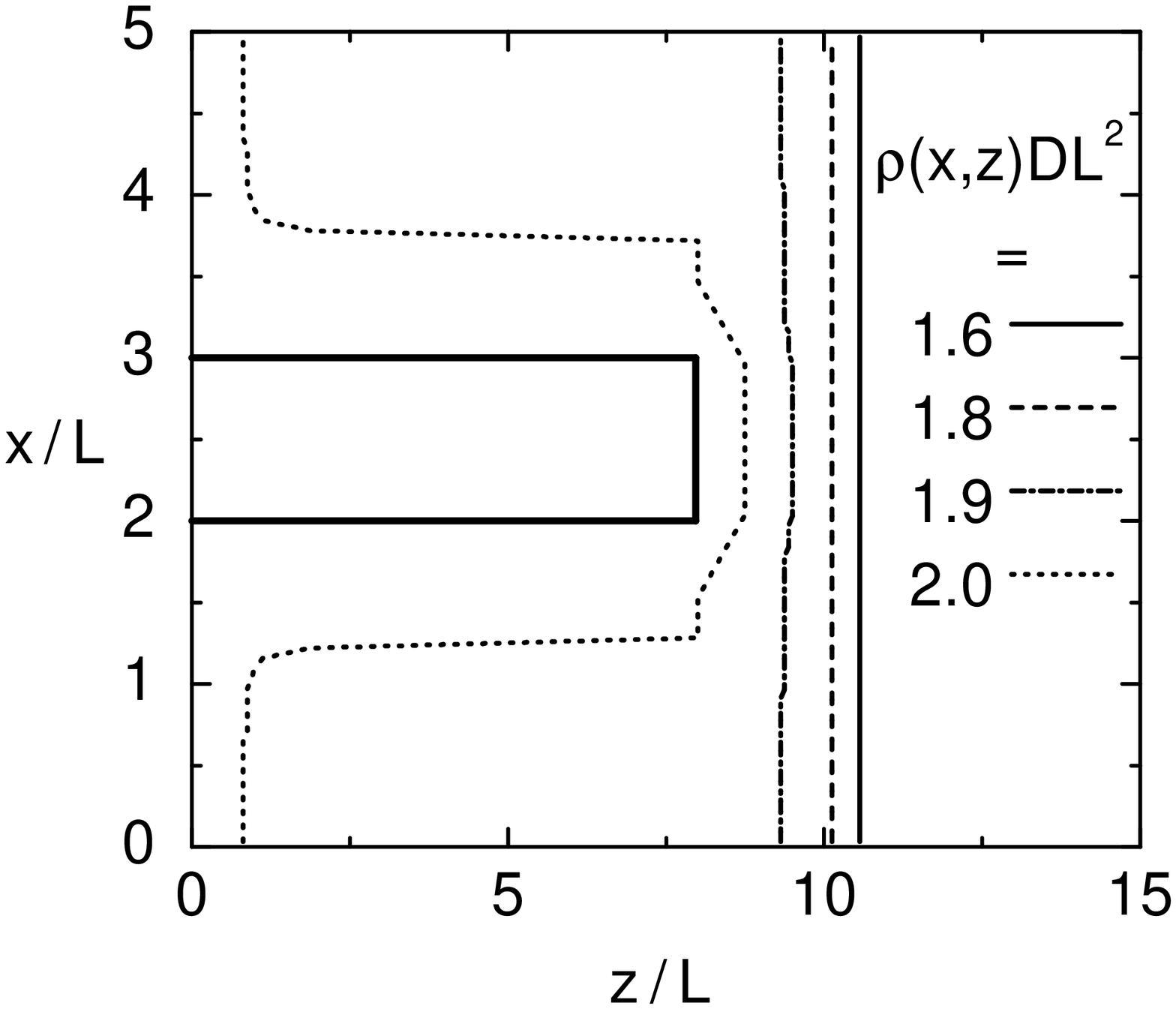}}
\caption{Orientationally averaged total density profile $\rho(x,z)$ 
[see Eq.~(\ref{eq11})] of thin rods of length $L$ and diameter $D$ 
($D/L \to 0$) in contact with a hard wall for three chemical 
potentials $\mu=0.75$, $0.77$, $0.808$ $k_BT$, where 
$\mu^{(IN)}/(k_BT)=0.8087227$ is the chemical potential at bulk 
isotropic -- nematic coexistence. The corresponding contour 
plots are shown on the right. The sharp structures in the contour lines
are caused by the discreteness of the orientational degrees of freedom.
The wall at $z=0$ is patterned with rectangular (parallel) barriers of 
width $w=L$ with $x_0=2.5\,L$, height $h=8\,L$, and periodicity 
$p=5\,L$ (see Fig.~\ref{fig1}). The corresponding excess adsorption 
is represented by the dash-dotted line in  Fig.~\ref{fig7} (b), and the 
three chemical potentials are marked by arrows in  Fig.~\ref{fig7} (b).
For $\mu=0.808\, k_BT$ the interface between the higher-density nematic film 
on the wall and the lower-density isotropic bulk fluid ($z\to\infty$) 
resembles closely the interface between the free isotropic -- nematic interface 
between coexisting bulk phases, with coexisting densities 
$\rho_b^{(I)}DL^2=1.25822486$ and $\rho_b^{(N)}DL^2=1.91544377$.
At $\mu\simeq 0.757$ $k_BT$ the trenches undergo a nematization filling
which is a smooth but steep variation of the density distribution.}
\label{fig8}
\end{figure}

To understand the origin of the calculated excess adsorptions, it is instructive 
to study the variation of the density profiles with increasing chemical potential.
The orientationally averaged density profiles $\rho(x,z)$ shown in Fig.~\ref{fig8} 
demonstrate that the wetting of the non-planar wall -- isotropic liquid interface by 
a higher-density nematic film occurs as a two-stage process where first the nematic 
phase  fills the space between the barriers until an almost planar 
isotropic -- nematic liquid interface has formed separating the higher-density nematic 
fluid in the space between the barriers from the lower-density isotropic bulk fluid. 
In the second stage a nematic film of diverging film thickness develops upon 
approaching the chemical potential at bulk isotropic -- nematic coexistence. In the 
presence of the patterned wall, the director (the average orientation of the 
rods) of the nematic phase is parallel to the $y$-axis because of the 
aforementioned effective entropic attraction between rods oriented parallel to the 
right-angled wedges of the barriers [see Fig.~\ref{fig6} (c)]. 
Upon approaching 
the chemical potential at bulk isotropic -- nematic coexistence, i.e., $\mu>0.757 \,k_BT$
in Fig.~\ref{fig7} (with this value of the kink position being largely independent of $h$), the calculated 
density profiles at the isotropic -- nematic interface become virtually indistinguishable 
from the free isotropic -- nematic interface between coexisting bulk phases, as expected 
for the case of complete wetting. In this limit this holds irrespective of the 
actual values of the width $w$, height $h$, and periodicity $p$. However, the 
thickness of the emerging nematic film depends on the aforementioned model parameters 
and the chemical potential. 

\begin{figure}[t]
\vspace*{-2cm}
\includegraphics[width=8.5cm]{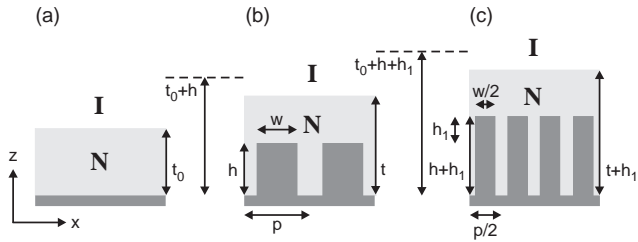}
\vspace*{0.5cm}
\caption{Schematic side view of a planar wall and two geometrically patterned 
walls (dark grey regions). The $y$-direction is perpendicular to the plane 
of the figure (see Fig.~\ref{fig1}). (a) The planar wall -- isotropic (I) liquid 
interface is wetted by a nematic (N) film of thickness $t_0$. 
(b) At the same chemical potential the film 
thickness $t$ on the geometrically patterned wall is larger than $h$ and 
smaller than $t_0+h$, where $t=t_0+h$ for $w=p$. 
(c) Illustration of Eqs.~(\ref{eq15}) and (\ref{eq16}); $w$ and $p$ are 
half as large as in (b) and $h$ is increased by $h_1$.}
\label{fig9}
\end{figure}

A very {\it thin} nematic liquid layer -- corresponding to a large undersaturation -- 
follows the substrate 
pattern, whereas a sufficiently thick layer is essentially flat (see Fig.~\ref{fig8}). 
In the following, we will be exclusively concerned with {\it thick} nematic films, 
allowing us to define an $x$-independent film thickness 
$t=L\times f(\frac{h}{L},\frac{p}{L},\frac{w}{L},\frac{\mu}{k_BT})$ defined as the distance 
$z_m$ between the midpoint of the density profile for the planar isotropic -- nematic 
interface and the substrate surface $z=0$; $f$ is a scaling function appearing on 
the basis of dimensional analysis. The minimal film thickness for which a planar 
isotropic -- nematic 
liquid interface is still possible is $t\gtrsim h$. We consider the situation 
as illustrated in Fig.~\ref{fig9} (a), where for a given chemical potential $\mu$ 
a nematic (N) film of thickness $t_0$ intrudes between a planar hard wall and an 
isotropic (I) bulk fluid. At the same chemical potential the film thickness $t$ of the fluid 
in contact with a geometrically patterned wall is larger than the height of the 
barriers $h$ and smaller than $t_0+h$ (see Fig.~\ref{fig9} (b)). For $p=w$
it follows that $t=t_0+h$ as expected on physical grounds. It is worthwhile 
to mention that $f(\frac{h}{L},\frac{p}{L},\frac{w}{L},\frac{\mu}{k_BT})$ 
as function of $w/L$ for given $h/L$, $p/L$, and 
$\mu/(k_BT)$ exhibits a discontinuity upon approaching $w \to 0$ because the 
ratio of the actual substrate area (including the side planes of the rectangular blocks)
per period $p$ over the one projected onto the $x-y$ plane drops abruptly 
from $(p+2h)/p$ to $1$ for $w\equiv 0$. Hence, due to geometric constraints the film 
thickness in the presence of infinitely thin barriers differs from the one in the 
absence of the barriers. Moreover, we find the following properties, which are schematically 
visualized 
in Figs.~\ref{fig9} (b) and (c):
\begin{equation} \label{eq15}
t=L\times f(\frac{h}{L},\frac{p}{L},\frac{w}{L},\frac{\mu}{k_BT})=
L\times g(\frac{h}{L},\frac{w}{p},\frac{\mu}{k_BT})
\end{equation}
and for fixed $w/p$ and $\mu/(k_BT)$
\begin{equation} \label{eq16}
t(h+h_1)=h_1+t(h)\,,
\hspace{0.5cm} h\gtrsim L\,.
\end{equation}
Equation (\ref{eq15}) states that upon varying the barrier width $w$ and the periodicity 
$p$ such that the ratio $(p-w)/w$ of the substrate area at the bottom $z=0$ over the 
substrate area at the top $z=h$ is kept fixed, the film thickness does not change for a 
given barrier height $h$, chemical potential $\mu$, and $L$. This result is reminiscent 
of the Cassie equation \cite{cass:48}. Cassie considered a simple liquid in contact with 
a smooth but chemically striped surface with periodicity $p$ such that on stripes of 
width $w$ one has a contact angle $\vartheta_1$ and in between $\vartheta_2$. The apparent 
average contact angle is given by
\begin{equation} \label{eq17}
\cos\vartheta_{app}=\frac{w}{p}\cos\vartheta_1+\left(1-\frac{w}{p}\right)
\cos\vartheta_2\,.
\end{equation}
Since $\vartheta_{app}$, $\vartheta_1$, and $\vartheta_2$ are determined uniquely
by $t_{app}$, $t_1$, and $t_2$ via the corresponding effective interface potentials 
(see Eq.~(4.56) in Ref.~\cite{diet:88}), Eq.~(\ref{eq17}) states that the apparent 
film thickness $t_{app}$ depends only on the ratio $w/p$, which is the analogue 
of Eq.~(\ref{eq15}). 
Whereas Eq.~(\ref{eq15}) is valid for all values of $w/L$, $h/L$, 
and $p/L$, for molecular-scale surface patterns with barrier heights $h\lesssim L$ we 
find deviations from Eq.~(\ref{eq16}).

The  wetting of the non-planar wall -- isotropic 
liquid interface by a nematic film is driven by the steric interaction of the 
rods with the solid substrate which is mediated by the fluid occupying the space 
between the barriers. As the barrier height $h$ increases the interaction 
between rods located at $z>h$ and the part of wall which is located at $z=0$
weakens, such that for large height $h$ and small width $w$, wetting is dominated 
by the interaction with the {\it fluid} in the space between the barriers and not 
by that with the {\it solid} substrate at $z=0$. However, one has to take into account 
that the rods in the trenches between the barriers interact not only with the part of 
the substrate wall which is located at $z=0$ but also with the side planes of the 
rectangular blocks which are parallel to the $z-y$ plane.

\section{summary}
We have studied hard-rod fluids near geometrically structured walls using Zwanzig's 
model of square parallelepipeds with only three allowed orientations (Fig.~\ref{fig1}).
Within the framework of a density functional theory, the grand potential functional 
is minimized numerically and density profiles, excess adsorptions as well as surface 
and line tensions are determined leading to the following main results:

(1) The line tension for the isotropic fluid in contact with a right-angled wedge 
(see Fig.~\ref{fig2}) exhibits a change of sign with increasing chemical potential
while the line tension for the fluid in contact with a right-angled edge as well as
the wall -- isotropic fluid surface tension at a planar hard wall are positive
(Fig.~\ref{fig3}). 

(2) Figures \ref{fig4} and \ref{fig5} demonstrate an enrichment [depletion] of rods 
lying parallel and close to the corner of a right-angled wedge [edge]. On the basis 
of effective entropic forces between the rods and the walls (see Fig.~\ref{fig6}),
the results may be interpreted in terms of a repulsive barrier of an effective 
potential repelling a rod, which is oriented parallel to the corner of an edge, 
and approaches the edge sidewise, and practically preventing it from passing around 
the corner. The effective potential acting on a rod which is oriented parallel to the 
corner of a wedge is larger than the one close to a planar wall. Building on the 
effects demonstrated in Figs.~\ref{fig4}, \ref{fig5}, and \ref{fig6}, it seems possible 
to devise structures that create localized and directional entropic 
force fields for both natural and synthetic rodlike colloids.

(3) Coexistence between an isotropic and a capillary condensed nematic phase is 
observed for the fluid confined in a hard pore of square cross-section, 
provided the cross-section is sufficiently large. The density profile of the capillary 
condensed nematic phase is characterized by a nematic phase throughout the pore, whereas 
the density profile of the coexisting phase decays towards an isotropic phase in the 
middle of the pore. For the same fluid confined in a slit pore the confinement effects 
are weaker, i.e., in the slit pore one observes capillary nematization only at a higher 
chemical potential.

(4) From the calculated excess adsorptions (Fig.~\ref{fig7}) and density profiles
(Fig.~\ref{fig8}) of a fluid consisting of hard rods near the geometrically 
structured wall shown in Fig.~\ref{fig1}, we conclude that complete wetting of the 
non-planar wall -- isotropic liquid interface by a nematic film occurs 
as a two-stage process. In the first stage the nematic phase  fills the 
space between the barriers until an almost planar isotropic -- nematic liquid interface 
has formed separating the higher-density nematic fluid in the trenches between the 
barriers from the lower-density isotropic bulk fluid. In the second stage a 
nematic film of diverging film thickness develops upon approaching the chemical 
potential at bulk isotropic -- nematic coexistence. The film thickness, defined as 
the distance between the midpoint of the density profile for the almost planar 
isotropic -- nematic interface and the substrate bottom at $z=0$, is larger for the 
fluid  near the geometrically structured wall than the one for the fluid near a planar 
wall at the same chemical potential (Fig.~\ref{fig9}).

Finally, we note that phenomena which emerge from the contact of a rod fluid which 
is in its bulk in the nematic phase are also interesting because of the possibility 
to deliver external lateral structures deep into the bulk of the adjacent fluid
which offers a convenient means to image patterned surfaces. Density functional theory 
will allow one to study also such a system.

\end{document}